\documentclass[11pt]{article}
\usepackage{fullpage}
\usepackage{graphicx}
\usepackage{amssymb,amsmath}

\def\rRichClub{\Phi}
\def\kRichClub{\phi}

\begin{document}
\begin{center}
{\Large Ensembles of reference networks based on the rich--club structure for non--evolving networks}\\[4ex]
Raul J. Mondrag\'on$^1$ and Shi Zhou$^2$\\[3ex]
$^1$School  of Electronic Engineering and Computer Sciences, \\Queen Mary University of London\\ Mile End Road, London, E1 4NS, United Kingdom \\[3ex]
$^2$University College London, Adastral Park Campus\\  Ross Building, Adastral
Park,\\Ipswich, IP5 3RE, United Kingdom
\end{center}

\begin{abstract}
In networks the rich nodes are the subset of nodes with large numbers of links, or high degrees.
The rich nodes and the connectivity between themselves (\emph{rich--club connectivity}) tend to
dominate the organisation of network structure. Recently there has been a considerable effort to
characterise  and model the rich--club connectivity in a variety of complex networks. In this paper
we firstly clarify a number of terms: the rich-club \emph{coefficient} quantifies the density of
connectivity between a subset of rich nodes; the rich-club \emph{structure} is the rich-club
coefficient measured across the hierarchies of nodes; the rich-club \emph{phenomenon} refers to the
dynamic behaviour responsible for the formation of the rich--club connectivity in evolving
networks; and the rich-club \emph{ordering} discerns whether the connectivity between rich nodes in
a non-evolving network (i.e.~closed network or network snapshot) is higher than a reference network
obtained by network randomisation. We then evaluate a recently proposed null model which is based
on an ensemble of reference networks conserving the degree distribution of the original network. We
remark that one should not confuse the rich-club structure of a network  with the rich-club
ordering detected by the null model. We also demonstrate that the null model cannot identify the
dynamical mechanism that generates the rich--club connectivity. The main contribution of the paper
is that we introduce two new ensembles of reference networks based on the rich-club structure for
non--evolving networks. The first ensemble preserves the rich-club coefficient (as a function of
the rank of a node) of the original network. Members of the ensemble exhibit similar degree
distribution as the original network, and for assortative networks, similar assortative mixing as
well. We propose that this ensemble can be used to study networks where assortativeness is a
fundamental property, e.g.~to detect the community structure in social networks. Analysis on the
ensemble also provides a different way to interpret and model the evolution of social networks. The
second ensemble preserves both the degree distribution and the rich--club coefficient (as a
function of node degree). The reference networks in this ensemble have a similar structure as the
original network. We use them to quantify the correlation profile between the rich nodes and
pinpoint which links between the nodes are the backbone of network structure.
\end{abstract}

\section{Introduction}
Many social, economic, biological and technological networks contain a small set of nodes which have  large
numbers of links, the so--called rich nodes. In some  networks the rich nodes are tightly interconnected
between themselves, forming a rich--club~\cite{Zhou2004}. The rich--club is an oligarchy in that it dominates
the organisation of the whole network. In scale--free networks~\cite{Albert02} the connectivity between the
rich nodes plays an important role in network functionality, for example in the transmission of rumours in
social networks~\cite{Masuda06} or the efficient delivery of data packets in the Internet~\cite{Zhou04G}. The
density of connections between the rich nodes is quantified by the rich--club coefficient~\cite{Zhou2004}.
The rich--club coefficient and its
generalisations~\cite{Colizza06,mcauley-2007,Wutchy07,Zlatic08,Opsahl08,Serrano08a} have been proved to be a
useful measure for studying complex networks.

Many of the complex networks studied in the literature are single networks in the sense that each network's
structure is  not one of several but unique, for example the Internet. In this case we do not have the
equivalent to a physical law to verify whether a statistical measure obtained from a single network is
expected or extraordinary. Instead, a common technique to assess the significance of a property of a single
network is to use the statistical randomisation method~\cite{Manly97} to create a null model. The procedure
consists of using the observed network to generate an ensemble~\cite{Bianconi08} of reference networks via
randomisation.  The null model is then generated from this ensemble.

In this paper we study three ensembles of reference networks which are generated by different network
randomisation processes. The first ensemble has been widely studied. In one recent study it was used to
assess whether the connectivity between the rich nodes in a network is due to chance or an unknown mechanism.
We discuss the results and clarify the confusions arising from the study. We then introduce the other two
ensembles, which are based on the rich--club structure for non-evolving networks. We analyse their properties
and propose useful applications for the statistical physics study on complex networks. Our work not only
advances the understanding of the rich-club structure in networks but also provides new methods for studying
other properties related to the rich-club structure.

\section{Definitions}

\subsection{The rich--club coefficient and the rich-club structure}

Degree of a node, $k$, the number of links a node has. The rank $r$ of a node is its position in the list of
decreasing order of node degrees, i.e.~the best-connected node is ranked as $r=1$, the second best-connected
node is $r=2$ and so on. Rich nodes can be defined as nodes with large degrees or small ranks. The
\emph{density} of connections between the $r$ richest nodes is quantified by the rich--club
coefficient~\cite{Zhou2004}
\begin{equation}\label{eq:richRank}
\rRichClub(r)=\frac{2E_{\leqslant r} }{r(r-1)},
\end{equation}
where $E_{\leqslant r}$ is the number of links between the $r$ nodes and $r(r-1)/2$ is the maximum number  of
links that these nodes can share. If $\rRichClub(r)=0$  the nodes do not share any link at all, if
$\rRichClub(r)=1$ the nodes form a fully connected sub--graph, a clique.
As a function of degree, the rich--club coefficient can also be given as~\cite{Colizza06}
\begin{equation}
\label{eq:richDegree} \kRichClub(k)=\frac{2E_{\geqslant k}}{N_{\geqslant
k}(N_{\geqslant k} - 1)},
\end{equation}
where $N_{\geqslant k}$ is the number of nodes with degrees greater or equal to $k$ and $E_{\geqslant k}$ is
the number of links between the $N_{\geqslant k}$ nodes.

The rich--club coefficients $\rRichClub(r)$ and $\kRichClub(k)$ are related but they are not the same. The
rank gives a unique label to each node, and the degree can be used to group nodes into subsets. If $r_k^*$ is
the node with degree $k$ such that $r_k^*+1$ is the rank of the node with degree $k-1$, then $\kRichClub(k) =
\rRichClub(r_k^*)$. This is to say, $\{\kRichClub(k) | k=0, \ldots, k_{\text{max}} \}$ is a subset of
$\{\rRichClub(r) | r=1,\ldots,N\}$, where $k_{\text{max}}$ is the maximum degree in the network and $N$ is
the total number of nodes. Two networks can have the same  $\kRichClub(k)$ and the same degree distribution
$P(k)$  for all $k$, but different $\rRichClub(r)$.

Originally the term  rich--club was defined as the set of the richest nodes that are tightly interconnected.
This definition is vague as `tight' is a relative concept. Recently, Valverde and Sol\'e~\cite{Valverde07}
proposed a criteria to define the rich--club  and hence the rich nodes. The rich--club is defined by the
existence of a crossover at $k_c$ in $\kRichClub(k)$ (or $r_c$ in $\rRichClub(r)$) and this crossover
characterise the rich nodes.

The rich--club \emph{structure} refers to  the density of connections across the hierarchies of nodes. It is
given by the rich--club coefficient $\kRichClub(k)$  for all $k$ (or $\rRichClub(r)$ for all $r$). The
rich--club structure can be fully defined by the degree--degree distribution, $P(k,\,k')$, the probability
that an arbitrary link connects a node of degree $k$ with a node of degree $k'$~\cite{Colizza06},
\begin{equation}
\kRichClub(k) = \frac{ N \langle k \rangle \sum_{k'=k}^{k_{\text{max}}} \sum_{k''=k}^{k_{\text{max}}}
P(k',k'')}{(N\sum_{k'=k}^{k_{\text{max}}}P(k'))(N\sum_{k'=k}^{k_{\text{max}}}P(k')-1) },
\end{equation}
where $\langle k \rangle$ is the average degree. Conversely, we show in the following that  fixing the
rich--club structure will constrain the degree--degree distribution~\cite{Zhou07a,krapivsky08}.
From the definition of the rich--club coefficient $\kRichClub(k)$, the number of links that have at one end
a node with degree $k$ and at the other end a node with degree at least $k$ is
\begin{equation}
\label{eq:constraintOne} E_{k}=E_{\geqslant k}-E_{\geqslant k+1}=
\kRichClub(k)\frac{N_{\geqslant k}(N_{\geqslant k}-1)}{2} -
\kRichClub(k+1)\frac{N_{\geqslant k+1}(N_{\geqslant k+1}+1)}{2}.
\end{equation}
In terms of the conditional probability $P(k'|k)$ that a node with degree $k$ has a neighbour with degree
$k'$,
\begin{equation}
\label{eq:constraintTwo}
E_{k}=2NP(k) \left( \sum_{k'=k}^{k_{\text{max}}}P(k'|k) - \frac{1}{2} P(k+1|k) \right),
\end{equation}
where is $NP(k)$ is the number of nodes with degree $k$, and the term  $P(k'|k) - \frac{1}{2} P(k+1|k)$ gives
the proportion of these nodes that are connected to a node with degree at least  $k$. We know that
$P(k,\,k')$ and $P(k'|k)$  are related as $P(k'|k)=\langle k \rangle P(k,k')/(k P(k))$ where $\sum_{k'}
P(k'|k) = 1$. Hence, fixing $\kRichClub(k)$ for all $k$ constrains the possible values of $P(k|k')$ or
equivalently $P(k,k')$.

\subsection{The rich--club phenomenon and the rich--club ordering} The rich--club phenomenon~\cite{Zhou2004}
refers to the \emph{dynamic} behaviour in some evolving networks where, if a node becomes rich, it will tend
to connect with other rich nodes forming a rich--club or join an existing rich--club. An evolving network
model can introduce such dynamic mechanisms to reproduce the rich-club phenomenon and therefore generate a
network with similar rich--club structure as the real network~\cite{Bornholdt01,Bar04,Zhou2004,Zhou04F}. One
such mechanism was introduced in 1942 by Simon~\cite{Simon55}. Simon's model was based on the addition of new
nodes and the addition of new links between nodes that belong to the same class, where a class is the set of
nodes with the same degree. Bornholdt and Ebel~\cite{Bornholdt01} have pointed out that this growth mechanism
allows different growth rate for different classes of  nodes and hence it can create a rich--club clique.
Recently Krapivsky and Krioukov~\cite{krapivsky08} showed how the inclusion of the rich--club phenomenon in
evolving networks drastically constraints their structure.

The term rich--club phenomenon has been mixed with the term \emph{rich--club ordering} in the recent
study~\cite{Colizza06} which compared the rich-club coefficient of the original network against a reference
network. We suggest to use the term rich--club ordering to refer to the increase of rich-club coefficient in
comparison with the reference network. It is a \emph{static} property obtained for non--evolving networks
(closed networks) or a snapshot of evolving networks.

\section{The relative rich--club reference network}

A widely studied ensemble of reference networks are the maximal random networks generated by
randomly reshuffling link--pairs of the network under study~\cite{Maslov02, Maslov04}. The
intrinsic structure of the original network is taken into account by imposing the restriction that
the reshuffling process should not change the degree distribution.

Recently Colizza~{\em et al.}~\cite{Colizza06} used this ensemble as a null model to discern whether the
connections between the rich nodes in a network is due to chance or due to an ``organisational principle''.
To do so, they compared the original rich--club coefficient $\kRichClub(k)$ with the randomised rich--club
coefficient~$\kRichClub_{\text{ran}}(k)$  obtained from the maximal random networks in the ensemble.
Colizza~{\em et al.}~suggested that the ``normalised'' rich--club coefficient
$\rho_{\text{ran}}(k)=\kRichClub(k)/\kRichClub_{\text{ran}}(k)$ discounts the structural correlation imposed
by the finite--size effects~\cite{Boguna04}.  If $\rho_{\text{ran}}(k)>1$, it means there is an
organisational principle that leads to an increase in the density of connections between rich nodes in a more
pronounced way than in the null model, i.e.~the rich--club ordering.

We prefer to call $\rho_{\text{ran}}(k)$ the \emph{relative} rich--club coefficient. Since the
number of nodes with degree $k$ does not change by the randomisation procedure,
$\rho_{\text{ran}}(k)=E_{\geqslant k}/E_{\text{ran},\geqslant k}$  is the ratio of
original--network links to the reference--network links. Hence $\rho_{\text{ran}}(k)$ does not give
information about the \emph{density} of connections between rich nodes (which is measured by the
rich--club coefficient). This simple observation is relevant because there has been  some confusion
in the literature on what the rich--club coefficient and  $\rho_{\text{ran}}(k)$ are measuring. For
example Fig.~1 shows the Internet network at the Autonomous System level
(AS--Internet)~\cite{Faloutsos99,Pastor01} and for the scientific collaborations network in the
area of condensed matter physics (Collaborations--A)~\cite{Newman01a,Newman01b}. Figs.~1(a) and (c)
show the two networks' original $\kRichClub(k)$ in green colour and values of the randomised
$\kRichClub_{\text{ran}}(k)$ obtained from $10^3$ maximal random networks, where the frequency of a
particular value of $\kRichClub_{\text{ran}}(k)$ is labelled with different colours, from seldom
(0.1, red) to often (1.0, blue). The null model $\langle \rho_{\text{ran}}(k) \rangle$ is obtained
by averaging over all the maximal random networks  and hence corresponds to the `bluest' dots.
For the AS Internet, the observation that $\langle \kRichClub_{\text{ran}}(k) \rangle >
\kRichClub(k)$, i.e. $\langle \rho_{\text{ran}}(k) \rangle<1$, for almost all values of $k$,
suggests that the Internet does not have a rich--club ordering. This has created the
misinterpretation as stated in~\cite{Colizza06,Amaral06} that the rich nodes in the AS Internet
were not tightly interconnected with each other. However, Fig.~1(b) shows that the 20
best-connected nodes in the AS Internet are tightly interconnected between themselves. Even more,
the top seven best-connected nodes form a fully connected clique. For the Collaborations--A
network, the property $\langle \kRichClub_{\text{ran}}(k) \rangle < \kRichClub(k)$ insinuates that
the top scientists form tighter collaborations compared to the reference networks and this has been
interpreted in~\cite{Colizza06,Amaral06} as the rich nodes are tightly interconnected. However, the
fact is that the top 20 best-connected nodes are sparsely interconnected in both the original
network and the reference networks.

We remark that the rich-club ordering of a network is a relative property which is based on the
comparison with reference networks generated from the network itself. It should not be confused
with the rich-club structure which is measured and compared between different networks by the
rich-club coefficient. Furthermore, as pointed out in~\cite{Jiang08}, for both the AS--Internet and
Collaborations network, the range of $\kRichClub_{\text{ran}}(k)$ increases as $k$ increases, hence
any assertion based upon the relationship between $\langle \kRichClub_{\text{ran}}(k) \rangle$ and
$\kRichClub(k)$ should be statistically tested.

\begin{figure}
\label{fig:one}
\begin{center}
\includegraphics[width=10cm]{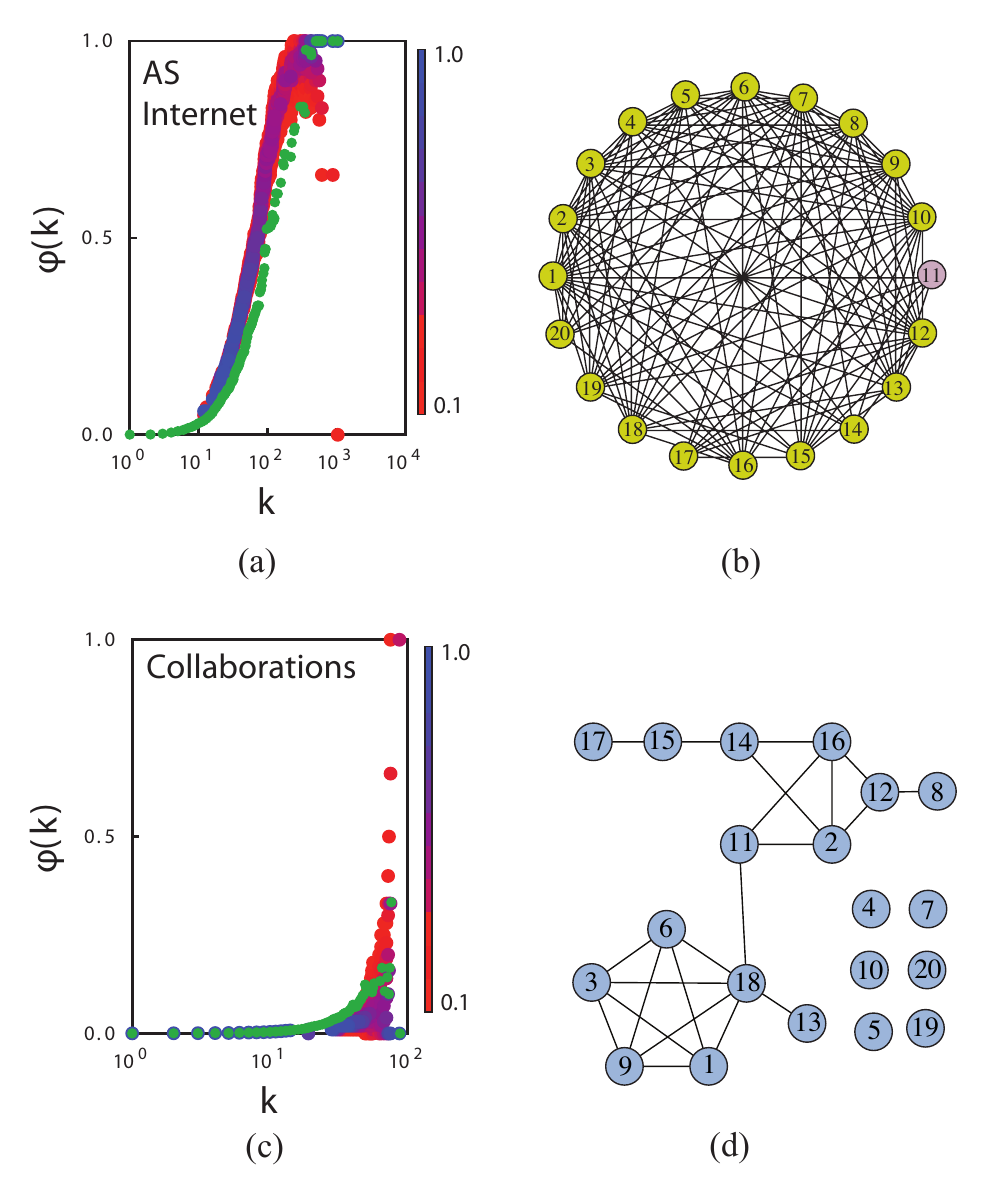}
\end{center}
\caption{ The rich--club coefficient $\kRichClub(k)$ (green) and  $\kRichClub_{\text{ran}}(k)$ (red
to blue) for the (a) AS--Internet and (c) the Collaborations network. For a given $k$, the range of
values of $\kRichClub_{\text{ran}}(k)$, obtained from $10^3$ maximal random networks, are divided
into 200 bins such that both the dispersion and the frequency of the values can be displayed in the
same graph. The frequency scale ranges from seldom (0.1, red) to often (1.0, blue). The null model
$\langle \rho_{\text{ran}}(k) \rangle$ corresponds to the `bluest' dots. (b) and (d) show the
interconnections between the 20 best-connected nodes in the two networks.}
\end{figure}

Here we show that the null model based on the maximal random networks cannot detect the rich-club
phenomenon, i.e.~whether there is a dynamic mechanism behind the formation of a rich--club.
Consider the preferential attachment mechanism introduced by Barab\'asi--Albert
(BA)~\cite{Barabasi99}. The preferential attachment correlates the age of a node with its
connectivity, i.e.~`rich--gets--richer'. As new nodes join the network the old nodes become richer.
However if two old nodes do not already share a link, they will never acquire a new one during the
network growth. This is to say, the BA growth mechanism is irrelevant to the formation of a
rich--club. Figure~2 shows that if a BA network grows from a fully connected seed, i.e.~a clique,
it will contain a fully connected rich--club; if it grows from a poorly connected seed, e.g.~a
ring, it will have a poorly connected rich--club.  The null  model $\langle
\kRichClub_{\text{ran}}(k) \rangle$ will produce contradictory results for the two networks
generated by the same BA growth mechanism. The null model in this case reflects the connectivity of
the initial seeds not the growth mechanism.

\begin{figure}
\label{fig:two}
\begin{center}
\includegraphics[width=13cm]{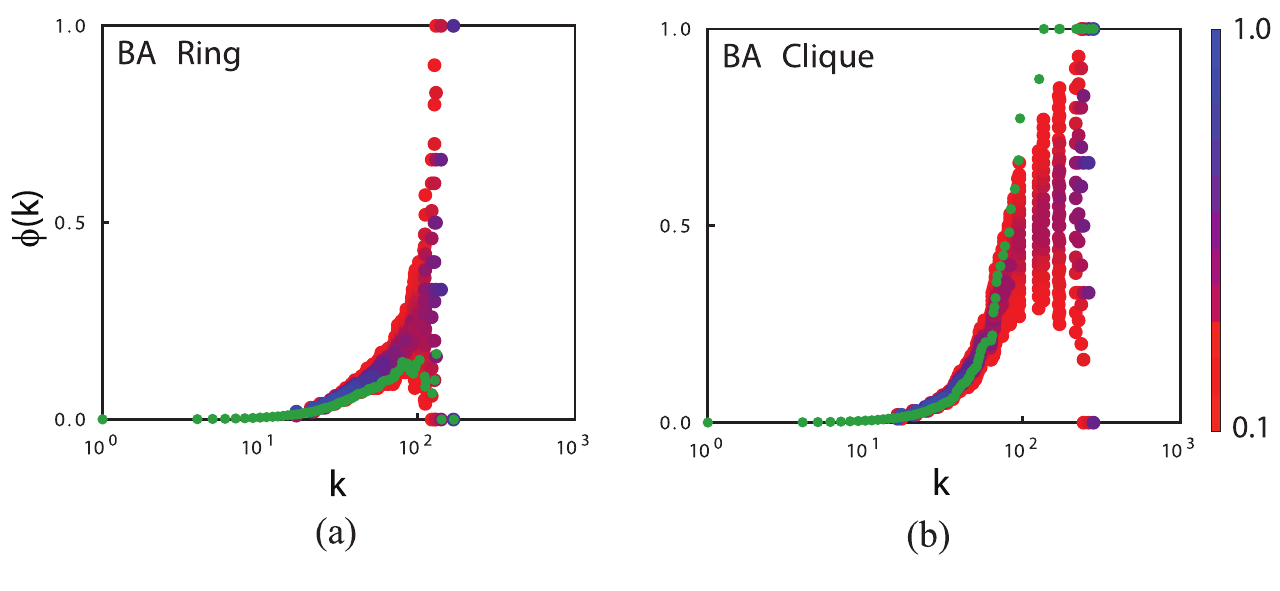}
\end{center}
\caption{ The rich--club coefficient for two Barab{\'a}si--Albert networks. Both networks have
$10^4$  nodes and are grown from network seeds of 10 nodes: one seed is a ring (BA--Ring) and the
other is a fully connected clique (BA--Clique). The colour scheme is the same as in Fig.~1.  }
\end{figure}

\section{The assortative reference network}

We can measure whether the maximal random networks used in the above null model discount the
degree-degree correlation in the original network by examining the nearest-neighbours average
degree of $k$-degree nodes~\cite{Pastor01}, $k_{\text{nn}}(k)=\sum_{k'=1}^{k_{max}} k' P(k'|k)$. If
$k_{\text{nn}}(k)$ is an increasing function of $k$, a network is assortative~\cite{Newman02}; and
if $k_{\text{nn}}(k)$ is a decreasing function of $k$, a network is disassortative. If
$k_{\text{nn}}(k)$ is neither  an increasing nor an decreasing function of $k$ then a network is
uncorrelated, more specifically, the degree of a node is independent of its neighbours' degree. If
the maximal random networks are less correlated than the original network, the slope of
$k_{\text{nn}}(k)$ should be less pronounced than the original network. Figure~3 compares
$k_{\text{nn}}(k)$  of the original network and the maximal random networks for four real networks:
the AS Internet, the protein interaction network of the yeast {\em Saccharomyces
cerevisiae}~\cite{Maslov02}, the Collaborations--A network, and the giant component of the
scientific collaboration network in the area of network theory~\cite{Newman06} (Collaborations--B).
Fig.~3 shows that the AS--Internet and the Protein networks are disassortative and their null
models are also disassortative, whereas the two collaborations networks are assortative and their
null models are uncorrelated networks.

\label{fig:three}
\begin{center}
\begin{figure}
\includegraphics[width=16cm]{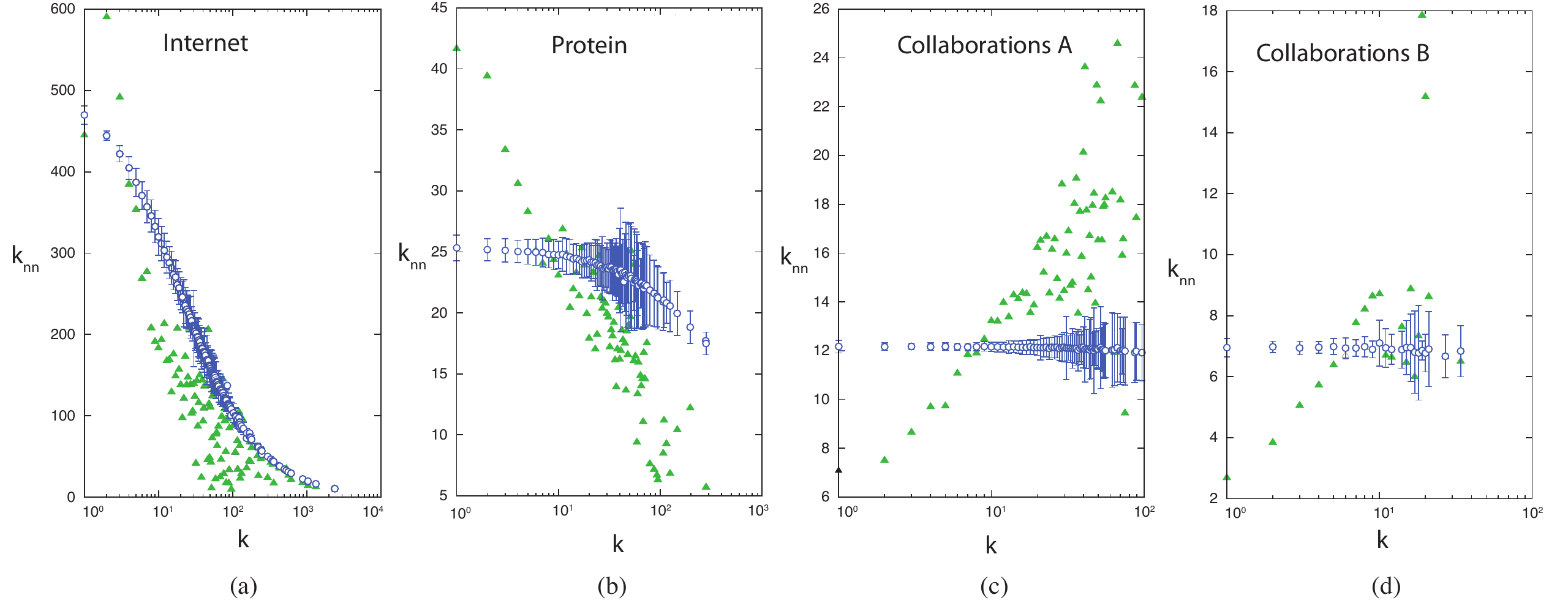}
\caption{Nearest-neighbours average degree of $k$-degree nodes, $k_{\text{nn}}(k)$, for the
original network (green) in comparison with the average and standard deviation of $
k_{\text{nn}}(k) $ for $10^3$ maximal random networks (blue). (a) The AS--Internet, (b) the
Protein, (c) the Collaborations--A and (d) the Collaborations--B networks.}
\end{figure}
\end{center}

The assortative mixing is the inherent property of social networks~\cite{Newman03}. Their maximal
random networks do not reflect this property. In the following we define a new ensemble of
reference networks, called assortative reference networks, which respect both the degree
distribution and the assortative mixing of social networks. We obtain such networks by
conserving\footnote{Note that conserving $\rRichClub(r)$ does not imply that the degree
distribution $P(k)$ and the rich-club coefficient $\kRichClub(k)$ are the same as in the original
network.} the original network's rich--club structure measured by the rich-club coefficient
$\rRichClub(r)$ as a function of rank $r$. We study the Collaborations--A network as a typical
example of assortative social networks.

\subsection{Progressive rewiring}

For a given network, we start with a random network having the same number of nodes and links as
the original network. A link is selected and rewired at random\footnote{We avoid creating
self-loop, duplicate link, or isolated node.}. We then evaluate the square error $\Delta =
\sum_{r=1}^N [\rRichClub(r)-\rRichClub^*(r)]^2$, where $\rRichClub(r)$ is the original rich-club
coefficient and $\rRichClub^*(r)$ is the rich-club coefficient of the rewired random network. If
the rewiring decreases the value of $\Delta$, then the rewiring is accepted; otherwise it is
rejected. This procedure continues until $\Delta$ is small.

\begin{figure}
\label{fig:ourNull}
\begin{center}
\includegraphics[height=18cm]{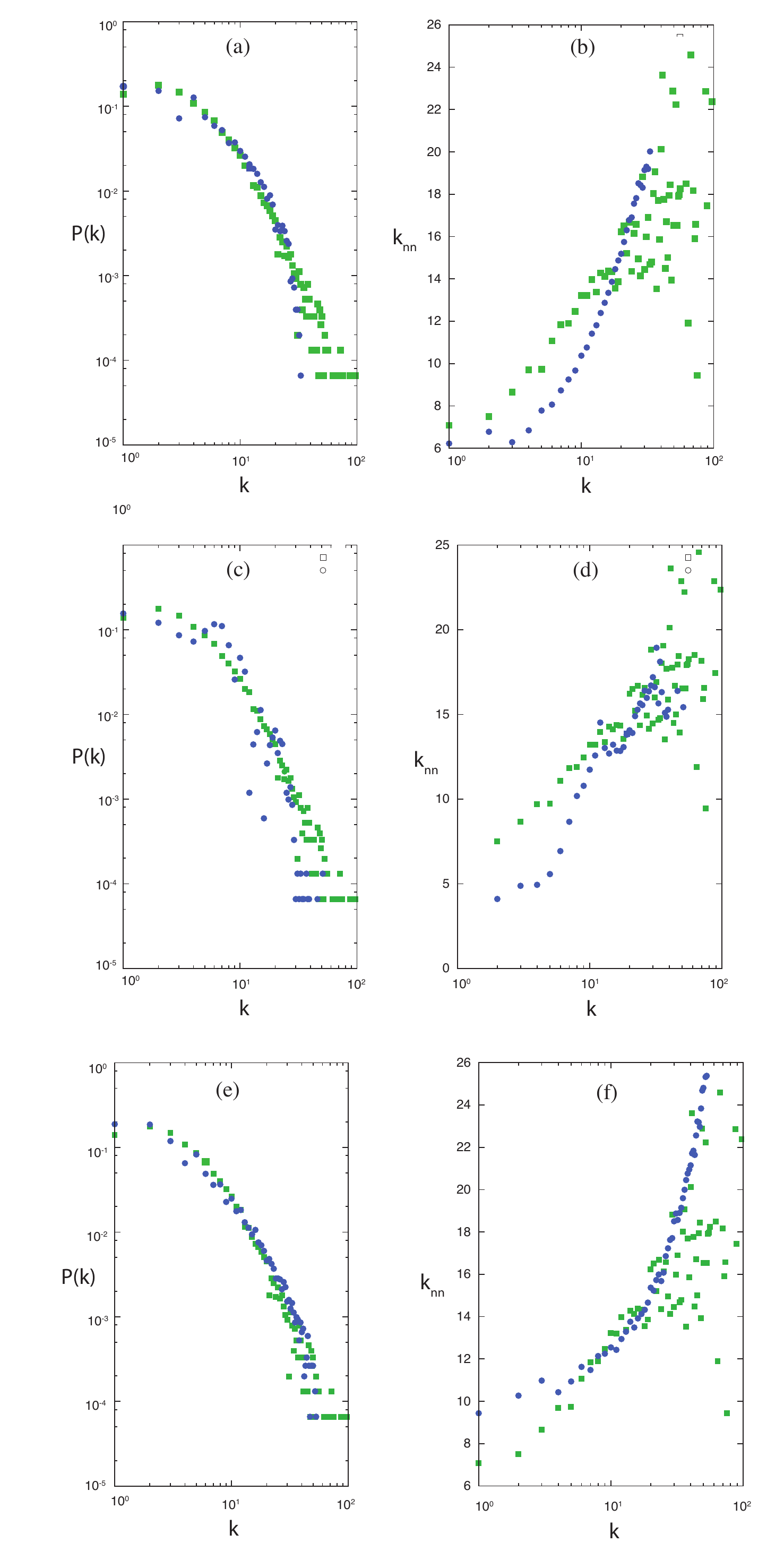}
\end{center}
\caption{Assortative reference networks for the Collaborations--A network. (a) (c) and (e) show the
degree distribution $P(k)$; and (b) (d) and (f) show the nearest-neighbours average degree of
$k$-degree nodes, $k_{\text{nn}}(k)$.  (a) and (b) are obtained from the progressive rewiring; (c)
and (d) are from the analytical solution; and (e) and (f) are from the rank-based preferential
attachment. The original network is shown in green and the reference network is in blue.}
\end{figure}

The assortative reference network obtained using this method not only conserve the rich-club
structure, but also resembles the degree distribution and the assortative mixing of the original
network (see Figures~4(a) and (b)). This is due to the structural constrain between the rich-club
structure and the degree-degree correlation (see Eqs.\,\ref{eq:constraintOne} and
\ref{eq:constraintTwo}). Basically, preserving $\rRichClub(r)$ of the $r$ richest nodes means
preserving the density of connections between nodes with the $r$ highest degrees.  Community
detection~\cite{Newman04} is based on the comparison between the density of connections of the
original network and reference networks. The assortative reference network could be useful for
detecting community structures of social networks, which are inherently assortative.

\subsection{Analytical solution}

Here we give the analytical solution for the assortative reference network. From the definition of the
rich--club coefficient $\rRichClub(r)$ (see Eq.\,\ref{eq:richRank}), the number of links that the node with
$r$ shares with the $r-1$ nodes of smaller ranks is
\begin{equation}
\label{eq:distdegrees} E_{r} = E_{\leqslant r}-E_{\leqslant r-1} =  \rRichClub(r)\frac{r(r-1)}{2} -
\rRichClub(r-1)\frac{(r-1)(r-2)}{2}.
\end{equation}
Assuming  the $E_{r}$  links are randomly  distributed between the $r-1$ nodes, the probability
that node $r$ has a link with node $r'$, where $r'<r$, is
\begin{equation}
\label{eq:distLinks} {\cal P}(r) = \frac{E_r}{r-1}.
\end{equation}
Thus the probability that there is a link between two nodes with ranks $i$ and $j$ is
\begin{equation}
\label{eq:probNull} p_{ij}=
\begin{cases}
{\cal P}(i), &  {\text{if }} i> j\cr {\cal P}(j), &  {\text{if }} i< j,\cr
\end{cases}
\end{equation}
The above equation satisfies the property that  $\sum_{i=1}^N\sum_{j=1}^N p_{ij} = 2L$,  where $L$
is  the total number of links of the original network. For a random network which has the same
number of nodes $N$ and the same rich-club structure $\rRichClub(r)$ as the original network, the
degree of node $i$ can be approximated by
\begin{equation}
\label{eq:estimation-k} k_i \approx \sum_{j=1}^N p_{ij},
\end{equation}
and the nearest-neighbours average degree of $k$-degree nodes is approximated by
\begin{equation}
\label{eq:estimation-knn}k_{\text{nn}}(k) \approx \frac{1}{N_k}\sum_{i; k_i=k}\left(\frac{1}{k}
\sum_{j=1}^{N} p_{ij} k_j\right),
\end{equation}
where the sum with index $i$  adds the nodes that have degree equal to $k$ and $N_k$ is the  number
of $k$-degree nodes.

Figures~4(c) and (d) confirms  that the analytical solution of the assortative reference network
has similar $P(k)$ and $k_{\text{nn}}(k)$ (obtained from  Eqs.\,\ref{eq:estimation-k} and
\ref{eq:estimation-knn}) as the original network. If the original network is assortative,
i.e.~high-degree nodes (with small $r$) tend to connect with high-degree nodes, we can expect from
Eq.~(\ref{eq:distLinks}) that the assortative reference network will also be assortative. However
if the original network is disassortative, i.e.~high-degree nodes tend to connect to low-degree
nodes (with large $r$), Eq.~(\ref{eq:distLinks}) will not capture this property as it does not
favour the connectivity between high-degree nodes and low-degree nodes.

\subsection*{Rank--based preferential attachment}
Eq.~(\ref{eq:probNull}) assumes a uniform probability for node $r$ to attach to any node $r'$ with
$r'<r$.  Here we modify Eq.~(\ref{eq:probNull}) to include a preferential attachment mechanism and
therefore favour the attachment to richer nodes with smaller $r$,
\begin{equation}
p_{ij}=
\begin{cases}
2j{\cal P}(i)/(i+1), &  {\text{if }} i> j\cr
2i{\cal P}(j)/(j+1), &  {\text{if }} i< j.\cr
\end{cases}
\end{equation}
This expression was obtained by using $\sum_{i=1}^N i = N(N+1)/2$ such that $2P(r)(r+1)\sum_{i=1}^r
i = E_r$. This is a preferential attachment based on the relative rank of nodes. Figures~4(e) and
(f) show that $P(k)$ and $k_{\text{nn}}(k)$ estimated from this rank--based preferential attachment
are very close to the original network.

This rank--based preferential attachment puts across a different way to study the collaborations
network in the context of its reference network. The original network is a member of an ensemble
which is defined by the density of connections between a referent group (rich nodes) and the total
number of nodes. If the probability of connection between two nodes is related to their rank
difference, then the collaborations network looks like a typical member of the ensemble. This
provides a different way to conjecture \emph{how} collaborations between scientists arise. In
general, a scientist will prefer to work with a scientist of the same or higher status, where
status is a relative concept\footnote{This does not explain \emph{why} the higher--status scientist
will agree to collaborate, perhaps to carry on more work, or to keep his/her high status.}.

\section{The reference network preserving $P(k)$ and $\phi(k)$}

Recently we introduced another ensemble of reference networks which are obtained by randomly
reshuffling link--pairs of the original network with the restriction that both $P(k)$ and $\phi(k)$
are preserved for all $k$~\cite{Zhou07a}. Consider a pair of links with end nodes $n_1$, $n_2$,
$n_3$ and $n_4$ with degrees $k_1<k_2<k_3<k_4$ respectively. If node $n_1$ is linked to $n_2$ and
$n_3$ is linked to $n_4$, we call they are assortatively wired. If a randomly chosen pair of links
are assortatively wired they are discarded and a new pair is considered; otherwise the four end
nodes of the links are reshuffled at random. The procedure is repeated  for a large number of
times.  Reshuffling a pair of links which are not assortatively wired  does not changes the number
of links between nodes with degrees equal or larger than $k$, hence $\kRichClub(k)$ is preserved.
Notice that this randomisation procedure also conserves the degree distribution $P(k)$. Fig.~5
shows that the original network and the reference  networks obtained from this method show similar
degree-degree correlation, and therefore have similar network structure.

\begin{figure}
\begin{center}
\label{fig:avKnnNullPandPhi}
\includegraphics[width=16cm]{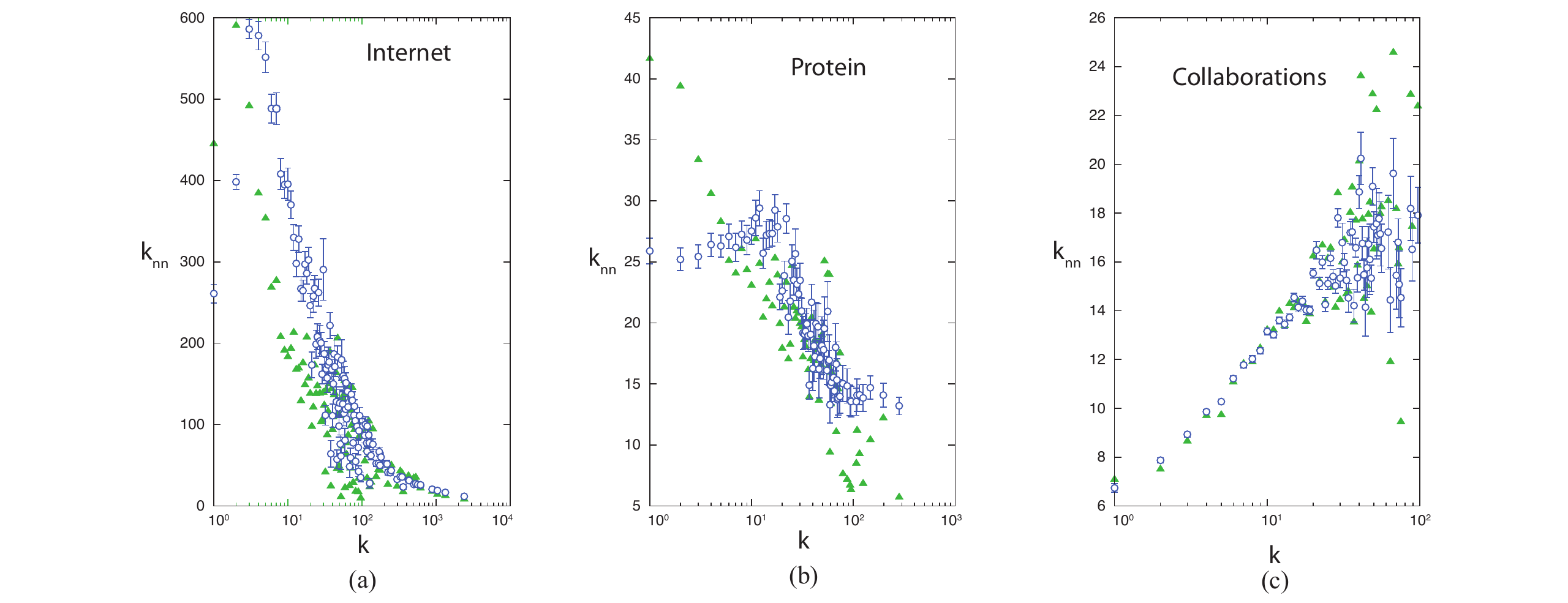}
\end{center}
\caption{The reference network preserving $P(k)$ and $\phi(k)$ for (a) the Internet, (b) the
Protein and (c) the Collaborations--A networks. Nearest-neighbours average degree
$k_{\text{nn}}(k)$ of the original network is in green and that of the reference networks is in
blue. For each network, $10^3$ reference networks are generated and averaged. The error bars give
the standard deviation for each value.}
\end{figure}

\subsection{Correlation profile between rich nodes}

The above ensemble of reference networks allows us to assess the correlation between rich nodes in
a network in relative to the randomised version of the network having similar network structure.
The correlation measures how important a connection between two rich nodes is in terms of the
organisation of the rich--club and the global structure.

Consider $a_{ij}$ as the $ij$-th entry of the adjacency matrix describing the original network, and
$a'_{ij}(m)$ is the $ij$-th entry of the adjacency matrix of the $m$-th reference network obtained
by the above randomisation process, where $M$ is the total number of the reference networks. The
frequency probability that there is a connection between nodes $i$ and $j$ in the $M$ reference
networks is
\begin{equation}
q_{ij}=\frac{1}{M}\sum_{m=1}^{M} a'_{ij}(m).
\end{equation}
The correlation profile is obtained by evaluating $b_{ij}=a_{ij}-q_{ij}$.  The case $b_{ij}=0$
happens if the connectivity of the original network and the reference networks is the same. This
case could happen if (1) the original network and the reference networks never have a link between
nodes $i$ and $j$; or (2) the original network and the reference networks always have a link
between nodes $i$ and $j$. The case $b_{ij}=1$ means the original network has a link between nodes
$i$ and $j$ but this link never appears in the reference networks. The case $b_{ij}=-1$ means that
in the original network there  is not a link between nodes $i$ and $j$ but this link always appears
in the reference networks.

\begin{figure}
\label{fig:correlation}
\begin{center}
\includegraphics[width=16cm]{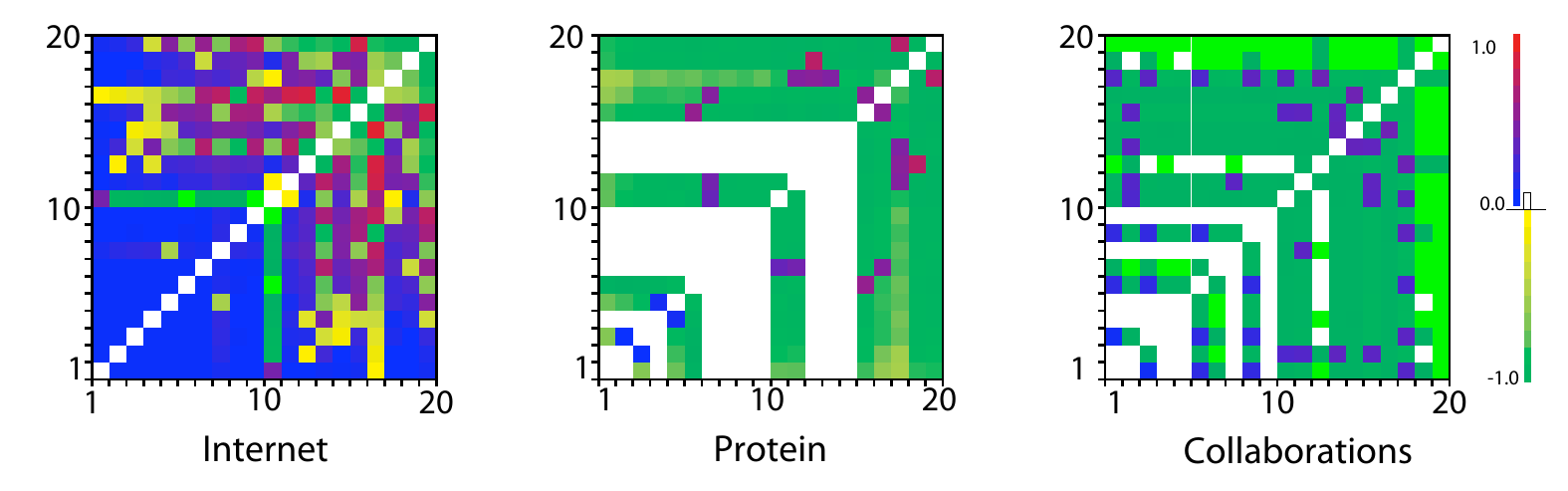}
\end{center}
\caption{ Correlation profiles for the top 20 richest nodes for the Internet, Protein and
Collaborations--A networks. For each network, the profile is obtained  from $10^3$ reference
networks preserving $P(k)$ and $\phi(k)$. The colour codes are: from blue to red ($b=0 \rightarrow
b=1$) labels the chance that a link exist in the original network but not in the reference
networks; from white to green ($b=0\rightarrow b=-1$) labels the chance that a link does not exist
in the original network but exist in the reference networks. }
\end{figure}

Figure~6 shows the correlation between the rich nodes in three real networks. The nodes are
labelled by their rank. As the case $b_{ij}=0$ can represent two different situations we label the
profile using two colour codes. The blue and white squares ($b_{ij}=0$) are the links that define
the backbone of the network. The existence, or not, of these links is fundamental for the network
structure and perhaps network functionality~\cite{Bianconi08}.

In the Internet profile,  the 7 richest nodes are always fully interconnected with each other in
the original network and in each of the reference networks (blue). Such a rich--club clique is an
important structure for the Internet because it provides a large number of shortcuts for the
delivery of data traffic and makes the average shortest path between any two nodes as small as just
over 3 hops\footnote{The AS-Internet contains more than 9 thousand nodes}. We conjecture that the
existence of such the rich--club clique is a fundamental property if the ensemble represents
networks that deliver traffic efficiently. Another interesting behaviour in the Internet's
correlation profile is that there is a link between nodes with ranks 15 and 17 that is present in
the original data but appears very rarely  in its randomised version (bright red square). Whether
this reflects erroneous or incomplete measurements, or for some reason these nodes interact with
each other against the odds, makes this interaction of particular interest.

The profile of the Protein network shows a completely different behaviour. The rich nodes in the
ensemble tend not to connect  with each other. The white vertical and horizontal bands show there
are interactions between proteins that never occur in the original network or the reference
networks. To decide if this structure is reflecting impossible protein interactions will require a
more specific analysis.

The profile of the Collaborations--A network  shows that some researchers tend to always
collaborate with each other (blue) and others never collaborate (white),  possibly reflecting the
friendship and rivalries between researchers. What is never present is a collaboration between two
researchers that it should not happen (no red squares in the profile), i.e.~hindsight cannot be
detected via the reference networks.

\section{Conclusions}
The rich nodes of a network tend to dominate the organisation of network structure so it is
relevant to understand if their interconnectivity is due to chance or to an organisational
principle. A technique to analyse this issue is to compare the rich--club coefficient against the
equivalent structure evaluated from an ensemble of reference networks. The ensemble is generated
via the restricted randomisation of the original network.

We evaluate a recently proposed null model which is based on the ensemble of maximal random
networks conserving the degree distribution of the original network. We remark that one should not
confuse the rich-club structure of a network  with the rich-club ordering detected by the null
model.

We presented a method to generate, from an assortative network, an ensemble of assortative
reference network with the condition that the rich-club coefficient $\rRichClub(r)$ is conserved.
The assortative reference network could be used for detecting the community structure of  social
networks which are inherently assortative. The assortative reference networks also  provides a
different way to explain the evolution of social collaborations.

We also presented a method to generate reference  networks which preserve both the degree
distribution and the rich-club coefficient $\phi(k)$. The ensemble of such reference networks have
a similar structure as the original network. We use them to study the correlations between the rich
nodes and pinpoint which connections between them form the backbone of the network.

Finally we'd like to remark that in the all cases of ensembles, the reference networks are
generated from the original network. Properties of the ensemble strongly depend on the properties
of the original network. Any deduction based on these ensembles should take into consideration this
dependance.

\section*{Acknowledgments}
RJM would like to thank the EPSRC~UK (EP/C520246/1) for support. SZ is supported by The Royal Academy of
Engineering and the UK Engineering and Physical Sciences Research Council (EPSRC) under grant no.~10216/70.

\bibliographystyle{plain}

\bibliography{RichClubNull}

\end{document}